\documentclass{PoS}
\newcommand{\be}{\begin{equation}}
\newcommand{\ee}{\end{equation}}
\newcommand{\br}{\begin{eqnarray}}
\newcommand{\er}{\end{eqnarray}}

\usepackage{graphicx}
\usepackage{subfigure}

\PoS{PoS(LAT2005)312}

\title{High accuracy simulations of d=4 SU(3) qcd-string }

\ShortTitle{High accuracy simulations of d=4 SU(3) qcd-string }

\author{\speaker{Nayasinganahalli Hari Dass}\thanks{
This work is part of the Xth plan project {\em Indian Lattice Gauge Theory Initiative} (ILGTI) at the Institute of Mathematical Sciences. We acknowledge
the generous support by the Department of Atomic Energy of India to this project.
}

        IMSc, Chennai\\

        E-mail: \email{dass@imsc.res.in}}

\author{Pushan Majumdar \thanks{Supported by Fonds zur F\"orderung der Wissenschaftlichen Forschung
in \"Osterreich, project M870-N08 (Lise-Meitner Fellowship)}\\

       Karl-Franzens Universit\"at Graz\\

        E-mail: \email{pushan.majumdar@uni-graz.at}}

\abstract{We present here results from our high accuracy simulations of
the $q\bar q$ potential in $d=4$ $SU(3)$ Yang-Mills theory. We measure
this quantity by measuring the Polyakov loop correlators using
the exponential variance reduction technique (multilevel) of L\"uscher
and Weisz. 
Further noise reduction was achieved by replacing temporal links
by their multihit averages. In this case we found that the 
semi-analytic 
 technique proposed by de Forcrand and
Roiesnel is much more efficient than the usual monte-carlo multihit. Measurements
were carried out on lattices of spatial extents of about 4 and 5.4 fermi. The temporal
extent was 5.4 fermi in both cases.
Polyakov loop correlators having separations between 0.3 and 1.2 fermi were measured. 
We analysed the results in terms of the force between $q\bar q$ pair
as well as in terms of a scaled second derivative of the potential.
The data is accurate enough to distinguish between different effective 
string models and it seems to favour the expression for ground state
energy of a Nambu-Goto string.

          }

\FullConference{XXIIIrd International Symposium on Lattice Field Theory\\

		 25-30 July 2005\\

		 Trinity College, Dublin, Ireland}

\begin{document}

\section{Introduction}
The mechanism of quark confinement continues to be a forefront problem
of theoretical physics despite decades of intense research both analytically
and through large scale numerical simulations. Lattice simulations strongly point 
towards the fact that confining flux tubes are indeed
formed between $q{\bar q}$ pairs. 
In recent times there has been intense interest
in the possible string-like behaviour of these flux tubes \cite{kuticonf}. Progress has been
significantly accelerated by the path breaking developments in algorithms like  
the exponential variance reduction techniques of L\"uscher and Weisz \cite{lw1}.
This has allowed one to measure Polyakov loop correlators to unprecedented
accuracy thereby increasing sensitivity to the sub-leading behaviour of the $q{\bar q}$  
potential.
In their work using the multilevel technique L\"uscher and Weisz directly measured,
 with high accuracy, the coefficient of the $1/r$
correction to the linearly rising part of the potential. For $d=3$ $SU(2)$ lattice gauge theories 
this had been measured long ago, though not to the same accuracy, by Ambjorn {\em et. al.} \cite{ambjorn}.
 Asymptotically, at large $r$ this coefficient is expected to have the universal value 
$c = -{(d-2)\pi\over 24}$ 
(called the L\"uscher term \cite{lterm})
characteristic of a large class of bosonic string theories. They found that at distances of about  
1fm the value of $c$ was still about 12\% away from this asymptotic value . L\"uscher and 
Weisz initially argued 
that even this
discrepancy could be accommodated through boundary terms in the string action \cite{lweisz} and 
that the
scale of string formation is around 0.5fm. Subsequently
they showed that the so called open string - closed string duality forbade such terms. Kuti
{\em et. al.} \cite{kuti} have undertaken very detailed studies of the spectrum of string excitations
for which they have used extended Wilson loops. According to them, the ordering of the spectrum 
 agrees with that of the bosonic string only at distances larger than 2fm. 
Other recent studies in lower dimensions and simpler gauge groups have been carried out in 
\cite{caselle,pushan}.

In this article we present results of our simulations for $d=4$ $SU(3)$ Yang-Mills theory and compare
with different string models.

\section{Our Simulations}

\subsection{Machine}

All simulations were carried out on the 
teraflop linux cluster KABRU built at IMSc, Chennai
for lattice simulations. The hardware configuration of this machine is: 144 dual Intel Xeons 
@ 2.4 GHz, 533 MHz FSB
and 266 MHz ECC DDRAM ( 2 GB per node on 120 nodes and 4 GB per node on 24 nodes). The networking
is done in 3-D torus topology with SCI technology from Dolphinics, Norway. The sustained
node to node bandwidth is 318 MB/s and latency is 3.8 microsecs between different nodes and 0.7 microsecs
on the same node. The sustained performance on HPL has been 1.002 teraflops (double precision). Scaling
on MILC codes has been very good : ks\_imp\_dyn1(75-80\%); on pure\_gauge(85\%). For more details see
\cite{kabru}.

\subsection{Algorithm}

Since we are interested in ground state properties, we measure Polyakov loop correlators 
 as it is well known that the Polyakov loop correlator has the best known projection
 onto the $q{\bar q}$
ground state. We use the 
multilevel technique of L\"uscher and Weisz for measuring the Polyakov loop correlators
very accurately. 
This algorithm crucially uses locality of the action and the consequent quasi-factorisation
of the partition function. 
%
%
%
%
%
%
For evaluating Polyakov loop correlators, the spatial-links on certain timeslices are held 
fixed first while all other links are updated. This is called sub-lattice updating. With each 
such update the direct-product of the product of temporal links in a sub-lattice at the locations 
of the Polyakov loop are evaluated and averaged over many sub-lattice updates . Finally these
direct-products are multiplied together and traced over to yield a Polyakov loop correlator.
This constitutes a single measurement. The final expectation value is an average over usually a 
few hundred such measurements.

For further details of the algorithm we refer the reader to \cite{lw1}.
    
This algorithm has several optimization parameters and the most important among them seems to be 
the number of sub-lattice updates used to compute the intermediate expectation values. We will refer to this
number henceforth as "iupd". Another parameter is the thickness of the sub-lattice.
For our case
a thickness of two was optimal.

It is also well known that improved observables can be constructed by replacing a link by its 
average as determined by the Boltzmann weight for the link. 
This is known as multihit and we employ that too on the time-like links for the 
correlators. The average is computed through the single-link integral defined by  
\be
\langle U\rangle = Z(J,J^\dag)^{-1}{\partial Z(J,J^\dag)\over\partial J^{\dagger}}\;\;\;\; 
{\rm where}\;\;\;\; Z(J,J^\dag) = \int_{SU(3)} [dU]e^{{\rm tr}(UJ^\dag+U^\dag J)}.
\ee
For SU(3) this average cannot be efficiently carried out analytically and is 
most often evaluated using a monte-carlo method. We used a semi-analytic method for this
averaging originally due to de Forcrand and Roiesnel \cite{forcrand}.
This resulted in a 60\% speedup of
the code compared to using the monte-carlo method to reach similar levels of accuracy.

\subsection{Simulation details}

We have carried out simulations at $\beta=5.7$ on both $24^3\times 32$ and $32^4$ lattices
using the Wilson gauge action.
The lattice spacing $a$
at this $\beta$ is 0.17 fm (as estimated through the Sommer scale) so that 
the temporal extent of the lattice is 5.4fm,
while the spatial box is $(4{\rm fm})^3$ in one case and $(5.4{\rm fm})^3$ in the other.

For Polyakov loop correlators 
separated by $r = 2-6$ each measurement involved simulations 
on $24^3\times 32$ 
lattices with iupd=12000 
and 500 measurements were made.
Errors were determined by jack-knife analysis. 
For the larger separations $r = 5-7$ simulations were done on $32^4$ lattices with 
iupd=24000 and in all 270
measurements were obtained.

Here it may not be out of place to mention 
that for the larger separations ($r=7$ currently), 
we had to go to larger lattices to continue to gain from the multilevel scheme. We observed 
that on a $24^3\times 32$ lattice increasing iupd even by an order of magnitude did not seem 
to help reduce the error 
on the $r=7$ correlator. However going to a larger lattice ($32^4$) did help significantly. At 
present we do not understand this very clearly, but we suspect finite volume effects in the 
sub-lattice expectation values might be the reason. 

\subsection{Some string actions and corresponding potentials}

\begin{figure}[!t]
\begin{center}
\includegraphics[width=0.4\textwidth,angle=-90]{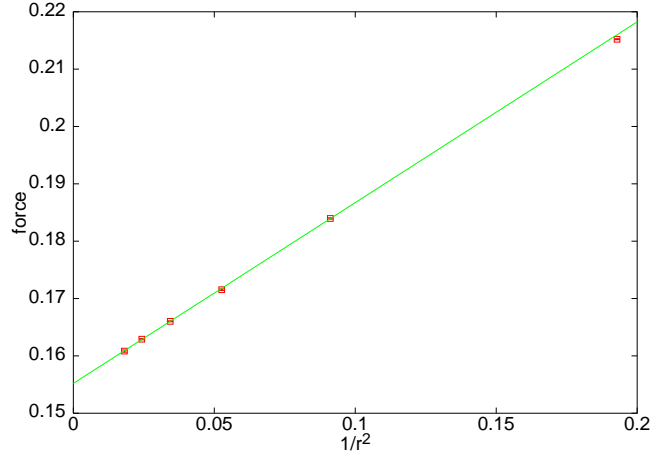}
\caption{Force in d=4 SU(3) case.}
\label{fig1}
\end{center}
\end{figure}

Before presenting our results we briefly recall 
 some string actions and
their ground state energies when their end-points are held fixed.

{\bf Case (a): free bosonic string}
\be
\label{free}
S = {{\sigma}\over 2}\int_0^T\int_0^R~~dz_0dz_1~~\partial_aX_\mu\partial_aX^\mu \;\;\;\;\Rightarrow \;\;\;\;
V(r) = {\sigma}r  - {{(d-2)\over 24}}{\pi\over r}
\ee
\indent
{\bf Case (b): Nambu-Goto string}
\be
\label{goto}
S = \sigma\int\int~~\sqrt{\det g}\;\;\;\;\Rightarrow \;\;\;\; V_{\rm Arvis}(r) = \sqrt{{\sigma}^2r^2-
{(d-2)\pi\over 12}\sigma}
\ee
The potential in the latter case was first given by \cite{arvis} and it should be noted that
it becomes purely imaginary for $r < r_c$ where $r_c = \sqrt{{(d-2)\pi\over 12\sigma}}$.
This  corresponds to the tachyon instability of the Nambu-Goto string.
The coefficient of the linear term of the potential ($\sigma $) is known as the string tension.

\subsection{Results and analysis}

In our simulations we measure the Polyakov loop correlator $\langle P^*P\rangle(r)$
where $r$ is the separation between the two Polyakov loops. The $q{\bar q}$ potential 
is determined as 
\be
V(r)=-\frac{1}{T}\log\langle P^*P\rangle(r)
\ee
where $T$ is the temporal extent of the lattice.
In our analysis we look at the force between the $q{\bar q}$-pair
given by $F(r) = -{\partial V(r)\over\partial r}$ and the scaled second derivative $c(r)$
given by $c(r) ={r^3\over 2}{\partial^2 V(r)\over \partial r^2}$. 
The latter takes the 
value ${- {(d-2)\pi
\over 24}}$ for case (a).
The dependence of $c(r)$ on distance distinguishes between different effective string models.
On the lattice we define these quantitites by 
\br
F({\bar r})&=&V(r)-V(r-1) \\
c({\tilde r})&=&\frac{{\tilde r}^{~3}}{2}[V(r+1)+V(r-1)-2V(r)]
\er
where ${\bar r}$ and ${\tilde r}$ are defined as in \cite{lw2} to reduce lattice artifacts.

Since different $r$ values 
are measured in the same simulation, they are quite strongly correlated.
Thus when one evaluates the differences in potentials, it helps to reduce errors 
if one 
evaluates the differences for each measurement and then averages over different 
measurements rather than the other way round. In practice we compute the 
difference between jackknife bins and then compute the jackknife error 
for the differences. 

In Fig.1 we have plotted the force $F(r)$ as a function of $1/r^2$. Fitting the data to the form $px+q$
with $p$ and $q$ as fit parameters, we obtain the string tension $a^2\sigma=0.1552(2)$ from $q$. 
We compute the Sommer scale $r_0$ (defined by $r_0^{~2}F(r_0)=1.65$) to be $2.93a$ from the force data.
\begin{figure}[!t]
\centering
\includegraphics[width=.6\textwidth,angle=0]{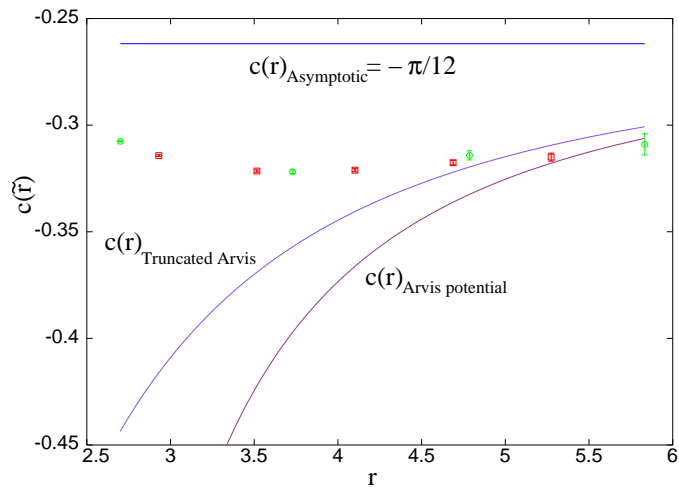}
\caption{Scaled second derivative for d=4 SU(3) case. The red points are from \cite{lw2}}
\label{fig2}
\end{figure}
In Fig.2 we have plotted the scaled double derivative $c({\tilde r})$ as a function of $r$. The horizontal
line at $c(r) = -\pi/12$ is the prediction of case (a). 
The pink curve is $c(r)$ as computed from the Arvis potential (case (b)) while the purple curve is $c(r)$
computed from the so called truncated Arvis potential which is nothing but a series expansion 
of the Arvis potential 
truncated at the second order ($O(1/r^2)$). 
It is clear that in the entire $r$-range shown, case(a)
is clearly excluded by our data. The force data also rules out case (a) quite convincingly as 
the fit parameter $p= 0.308$ 
is quite far from the value $\pi/12$. It should be noted that Kuti et al \cite{kuti} claim that their
data fits well the $q\bar q$-potential of case(a) in the this region.
Below 0.75fm the data deviates from case (b) also. We interpret this to mean that there is no obvious
string behaviour even upto 0.75fm. But as $r$ approaches 1fm we see clear convergence to case (b).

\section{Conclusions and future directions}
From our data we therefore see clear convergence towards the potential predicted by Nambu-Goto
theory for $r > 0.75{\rm fm}$. Similar conclusions are reached in \cite{pushan} for the 
3-d SU(2) case as well as by 
\cite{caselle} for $r > T/2$ though for $r < T/2$ they report deviations from Nambu-Goto theory.
But as noted by
many, including Arvis, the quantisations that led to the potentials in eqns(\ref{free},\,\ref{goto})
are consistent in only $d=26$ and can not be used to analyse the qcd-strings in $d=3,4$. 

The work of Polchinski and Strominger \cite{polchinski} may offer a resolution to this.
According
to them the free bosonic string action is inconsistent in $d\ne 26$ and it can be made consistent
in all dimensions by modifying (in a d-dependent manner) both the action and conformal 
transformations. Miraculously (till we understand this better) the $1/r$ correction to $V(r)$ is
the same as in cases (a) and (b)! But it appears highly unlikely that their approach would
produce $1/r^2$ and higher order terms that are close to what case (b) demands; yet our present work as well as
the works of \cite{caselle,pushan} are suggesting that case (b) works very well even in $r$-ranges
which are not so extremely large as to make $1/r^2$ corrections negligible. It is extremely important
to see whether the convergence to the Nambu-Goto case seen in our data holds for larger values. 
These need to  be sorted out through more careful simulations. 

Going to larger $r$-values, larger lattices and larger $\beta$-values (closer to continuum) will require
even greater computer resources unless we are able to come up with more efficient algorithms. We
are currently working on these. We also plan to study the implications of our data for 
other varieties of strings like those with extrinsic curvature etc. We also plan to investigate the 
simpler $Z_3$-gauge theories to learn about the issues already discussed here as well as to learn about
 string interactions.


\begin{thebibliography}{99}

  \bibitem{kuticonf} J.~Kuti,
\emph{ Lattice QCD and string theory}, these proceedings.
  \bibitem{lw1} M.L\"uscher and P. Weisz, 
\emph{ Locality and exponential error reduction in numerical lattice gauge theory},
JHEP {\bf 0109} (2001) 010 [{\tt hep-lat/0108014}].
  \bibitem{ambjorn} J. Ambjorn, P. Olesen and C. Petersen,
  \emph{ Observation of a string in three-dimensional SU(2) lattice gauge theory},
  Phys. Lett. {\bf B142}:410 (1984).
  \bibitem{lw2} M.L\"uscher and P. Weisz, 
\emph{ Quark confinement and the bosonic string}, JHEP {\bf 0207} (2002) 049 [{\tt hep-lat/0207003}].
  \bibitem{lterm} M.L\"uscher, K. Symanzik and P. Weisz, 
\emph{ Anomalies of the free loop wave equation in the WKB approximation}, Nucl. Phys. {\bf B173} (1980) 365;\\
M. L\"uscher, \emph{ Symmetry breaking aspects of the roughening transition in gauge theories}, Nucl. Phys. {\bf B180} (1981) 317.
  \bibitem{lweisz} M.L\"uscher and P. Weisz,
\emph{ String excitation energies in SU(N) gauge theories beyond the free-string approximation},
JHEP {\bf 0407} (2004) 014 [{\tt hep-th/0406205}].
  \bibitem{kuti} J.Juge, J.Kuti and C. Morningstar, 
\emph{ Fine structure of the QCD string spectrum}, Phys. Rev. Lett. {\bf 90} (2003) 161601 [{\tt hep-lat/0207004}];\\
\emph{ QCD string formation and the Casimir energy}, [{\tt hep-lat/0401032}].
  \bibitem{caselle} Caselle, Hasenbush and Panero, 
\emph{ Comparing the Nambu-Goto string with LGT results}, JHEP {\bf 0503} (2005) 026 [{\tt hep-lat/0501027}].
  \bibitem{pushan} Pushan Majumdar, 
\emph{ The string spectrum from large wilson loops}, Nucl. Phys. {\bf B664} (2003) 213 [{\tt hep-lat/0211038}];\\ 
\emph{ Continuum limit of the spectrum of the hadronic string}, [{\tt hep-lat/0406037}].
  \bibitem{forcrand} P. de Forcrand and Roiesnel, 
\emph{ Refined methods for measuring large-distance correlations}, Phys. Lett. {\bf B151} (1985) 77.
  \bibitem{arvis} J.F.~Arvis, 
\emph{The exact $q\bar q$ potential in Nambu string theory}, Phys. Lett. {\bf B127} (1983) 106.
  \bibitem{polchinski} J.~Polchinski and A.~Strominger, 
\emph{ Effective string theory}, Phys. Rev. Lett. {\bf 67} (1991) 1681.
  \bibitem{kabru} \href{http://www.imsc.res.in/~kabru}{{Kabru}}

\end{thebibliography}
\end{document}